\documentclass{article}
\usepackage{amsfonts,amssymb, amsmath}
\usepackage[english]{babel}
\usepackage{graphicx}
\usepackage{float}

\newtheorem{ex}{Example}

\def\nn{\nonumber }
\def\bq{ \begin{equation}}
\def\eq{ \end{equation}}
\def\ben{ \begin{eqnarray}}
\def\en{ \end{eqnarray}}

\begin{document}

%%%%%%%%%%%% TITLE %%%%%%%%%%%%%%

\title{  Reduction of divisors and  Kowalevski top}

\author{A.V. Tsiganov \\
\it\small St. Petersburg State University, St. Petersburg, Russia\\
\it\small email: andrey.tsiganov@gmail.com}
\date{}
\maketitle

\begin{abstract}
In the modern theory of the Kowalevski top there are  two elliptic curves introduced by  Kowalevski and by Reyman and Semenov-Tian-Shansky.
The Kowalevski variables of separation and poles of the  Baker-Akhiezer function define two classes of linearly equivalent divisors on these elliptic curves.
According to the Riemann-Roch theorem each class  has a unique reduced representative and we construct  such reduced divisors for the Kowalevski top.
 \end{abstract}

\section{Introduction}
\setcounter{equation}{0}
In the realm of algebraic geometry usually associated with many
Liouville integrable systems, the Hamiltonian evolution equations are written as a Lax equation
\[
\frac{d}{dt} L(x)=[L(x),A(x)]\,,
\]
for two $N\times N$ matrix functions  $L(x)$ and $A(x)$ on the phase space depending on  the auxiliary spectral parameter $x$.
The time-independent spectral equation
\bq\label{ba-f}
L(x)\, \psi(x,y) = y\, \psi(x,y)
\eq
allows us to represent the vector  Baker-Akhiezer function $\psi$ in terms of the Riemann theta function on a nonsingular compactification of the spectral curve defined by the equation
\bq\label{s-curve}
\Gamma:\qquad f(x,y)=\det(L(x)- y)=0\,.
\eq
There is a large freedom of similarity transformations of the Lax matrix,
\[L(x) \to  V L(x) V^{-1}\,,\]
which do not change the spectrum of $L(x)$, but change poles of $\psi$ which form a  $D$  on $\Gamma$. This
freedom can be characterized, and therefore fixed, by introducing a normalization of the
Baker-Akhiezer function
\[
\vec{\alpha} \cdot \psi=\sum_{i=1}^N \alpha_i\psi_i=1\,,
\]
which is given by a normalization (row-) vector $\vec{\alpha} = (\alpha_1,\cdots,\alpha_N)$ \cite{skl95}.

Any divisor $D$ determines a class of  linear  equivalent divisors
\[
D\to \{D,D_1,D_2,\ldots, D_k,\ldots\}\,,\qquad D-D_k=div(f_k)\,,
\]
where $div(f_k)$ is  divisor of the rational function $f_k$ on  $\Gamma$ \cite{mir95,mum84}.
It follows from the Riemann-Roch theorem that each class  has a unique
reduced representative  $\rho(D)$. Similarly, any normalization $\vec{\alpha}$ determines  a family of equivalent normalizations
\[
\vec{\alpha}\to\{\vec{\alpha}\,,\vec{\alpha}_1\,,\vec{\alpha}_2\,,\ldots,\vec{\alpha}_k,\ldots\}
\]
associated and unique normalization $\rho(\vec\alpha)$ associated with  $\rho(D)$, see examples in \cite{ts20}.

In this note, we construct  two reduced divisors using  Kowalevski's  variables of separation and  poles of the Baker-Akhiezer function
 which is an eigenfunction of the $4\times 4$ Lax matrix proposed by Reyman and  Semenov-Tian-Shansky \cite{rs87}. According to Kuznetsov \cite{kuz022} these
poles are also variables of separation in a particular case of the Kowalevski top which is contrary to the Dubrovin-Skrypnyk theory \cite{du98} that does not take into account 
reducibility of abelian varieties.

\subsection{Description of the model}
Let two vectors $\ell$ and $g$ are coordinates on
the phase space $M$. As a Poisson manifold $M$ is identified
with Euclidean algebra $e(3)^*$ with the Lie-Poisson brackets
\begin{equation}\label{e3}
\,\qquad \bigl\{\ell_i\,,\ell_j\,\bigr\}=\varepsilon_{ijk}\ell_k\,, \qquad
\bigl\{\ell_i\,,g_j\,\bigr\}=\varepsilon_{ijk}g_k \,, \qquad
\bigl\{g_i\,,g_j\,\bigr\}=0\,,
\end{equation}
 having two Casimir functions
\bq\label{caz-fun}
c_1=g_1^2+g_2^2+g_3^2\,,\qquad c_2=g_1\ell_1+g_2\ell_2+g_3\ell_3\,.
\eq
Here  $\varepsilon_{ijk}$ is the skew-symmetric tensor.

The Euler-Poisson equations on $e(3)^*$ are given by
\begin{equation}\label{eqm-g}
\
 X:\qquad \dot{\ell}=\ell \times\frac{\partial H}{\partial \ell}
 +g\times\frac{\partial H}{\partial g},\qquad
\dot{g}=g\times \frac{\partial H}{\partial \ell}\,,
\end{equation}
where $x\times y$ means a cross product of two vectors.

The Kowalevski top is defined by the Hamiltonian $H$,
\begin{equation}\label{h}
H=\ell_1^2+\ell_2^2+2\ell_3^2-2bg_1,\qquad {b}\in\mathbb R
\end{equation}
and the second integral $K$,
\begin{equation}\label{k}
K= (\ell_1^2 + \ell_2^2)^2+4b \bigl(g_1(\ell_1^2-\ell_2^2) + 2g_2\ell_1\ell_2\bigr)+4b^2(g_1^2 + g_2^2)\,,
\end{equation}
which are in  involution  $\{H,K\}=0$ with respect to the Poisson brackets (\ref{e3}).

\section{Reduced divisor on the Kowalevski elliptic curve}
\setcounter{equation}{0}
The Euler-Poisson equations (\ref{eqm-g}) were integrated by  Kowalevski by using  change of variables
which reduced the problem to hyperelliptic quadratures \cite{kow89}. Let us briefly discuss her calculations.

At the first step, Kowalevski introduced  two pairs of Lagrangian variables $z_{1,2}$ and $\dot{z}_{1,2}$ such that
\bq\label{hk}
 H=
-\dfrac{\dot{z}_1\dot{z}_2+R(z_1,z_2)}{(z_1-z_2)^2}\,,\qquad
K=\frac{\bigl(\dot{z}_1^2-R(z_1,z_1)\bigr)\bigl(\dot{z}_2^2-R(z_2,z_2)\bigr)}{(z_1-z_2)^4}\,.
\eq
Here
\bq\label{z12}
z_1=\ell_1 + {\mathrm i} \ell_2, \qquad z_2=\ell_1-{\mathrm i} \ell_2
\eq
and
\[
\label{R12} R(z_{1},z_{2} )=
 z_{1} ^2 z_{2} ^2-H(z_{1} ^2+z_{2} ^2) -4b\,c_2(z_{1} +z_{2} )-4b^2\,c_1+K.
\]
To remove cross-terms $\dot{z}_1\dot{z}_2$ in Hamiltonian  (\ref{hk}) Kowalevski used arithmetic of divisors
\bq\label{rot}
P'_1=P_1+P_2\,,\qquad P'_2=P_1-P_2\,,
\eq
on the elliptic curve  $E$ defined by equation
\bq\label{ell-1}
E:\qquad Z^2= R(z,z)\,,\qquad R(z,z)=z^4-2Hz^2 -8b\,c_2z-4b^2\,c_1+K\equiv \sum_{k=0}^4 a_kz^k.
\eq
Here  $P_k=(z_k,Z_k)$ and $P'_k=(z'_k,Z'_k)$ are two pairs of points defining two semi-reduced divisors $D=P_1+P_2$ and $D'=P'_1+P'_2$ on the abelian variety  $E$.
Transformation (\ref{rot}) can be rewritten in the matrix form
\[
\left(
              \begin{array}{c}
                P_1 \\
                P_2 \\
              \end{array}
            \right)\to \left(
              \begin{array}{c}
                P'_1 \\
                P'_2 \\
              \end{array}
            \right)=\left(
                            \begin{array}{cc}
                              1 & 1 \\
                              1 &-1 \\
                            \end{array}
                          \right)\left( \begin{array}{c}
                P_1 \\
                P_2 \\
              \end{array}
            \right)
\]
similar to the standard rotation which reduces quadratic form to a diagonal form.

According to Abel \cite{ab}, affine coordinates of points $P'_1=(z'_1,Z'_1)$ and $ P'_2=(z'_2,Z'_2)$  on $E$ are equal to
\bq\label{z-var}
z'_{1,2}=-z_1-z_2-\dfrac{2b_0b_2+b_1^2-a_2}{2b_1b_2-a_3}\qquad\mbox{and}\qquad Z'_{1,2}=-\mathcal V_{1,2}(z'_{1,2})\,.
\eq
Here $a_j$ are  given by (\ref{ell-1})  and $b_j$ are coefficients of the interpolation  polynomials
\[
\mathcal V_{1,2}(z)=b_2z^2+b_1z+b_0\equiv\sqrt{a_4}(z-z_1)(z-z_2)+\dfrac{(z-z_2)Z_1}{z_1-z_2}\pm\dfrac{(z-z_1)Z_2}{z_2-z_1}\,,
\]
which now are called second Mumford's coordinates of the semi-reduced divisors $P_1\pm P_2$, respectively.

Abscissas $z'_1$ and $z'_2$ commute to each other with respect to the Poisson brackets (\ref{e3})
\[\{z'_1,z'_2\}=0\]
and the corresponding velocities $\dot{z}'_1$ and $\dot{z}'_2$ satisfy to Abel's differential equations, see (\ref{abel-kow-z}) below.

A formal description of the group structure and arithmetic (\ref{z-var}) in a real elliptic curve
\[
y^2=a_4x^4+a_3x^3+a_2x^2+a_1x+a_0\,,\qquad a_4\neq 0
\]
is   a little more difficult than arithmetic in an imaginary elliptic curve
\[
y^2=a_3x^3+a_2x^2+a_1x+a_0\,,\qquad a_3\neq 0\,,
\]
because these curves differ by the number of points at infinity,  see \cite{pa99} and references within.

According to Weierstrass,   any real hyperelliptic  curve  with a ramified prime divisor can be reduced to
the imaginary hyperelliptic curve by a birational transformation. For these reasons
 Kowalevski reduced the real elliptic curve $E$ (\ref{ell-1}) to the cubic
\bq\label{ell-2}
E:\quad W^2=4 w^3+a w+b\,,
\eq
where
\[
a=4b^2c_1-K- \frac{H^2}{3}\,,\qquad b=\frac{H(36b^2c_1 + H^2 - 9K)}{27}-4b^2c_2^2\,,
\]
by using birational transformation from the unpublished Weierstrass lectures \cite{kow89}.

This birational transformation maps $z'_{1,2}$ to variables $w_{1,2}$ which  satisfy Abel's differential equations
\bq\label{w-ab-eq}
\frac{\dot{w}_1}{\sqrt{\mathcal R_5(w_1)}}+\frac{\dot{w}_2}{\sqrt{\mathcal R_5(w_2)}}=0\,,\qquad
\frac{w_1\dot{w}_1}{\sqrt{\mathcal R_5(w_1)}}+\frac{w_2\dot{w}_2}{\sqrt{\mathcal R_5(w_2)}}=1
\eq
on hyperelliptic curve
\[
C:\qquad y^2=\mathcal R_5(w)\,,
\]
where $\mathcal R_5(w)$ is the reducible polynomial
\[
\mathcal R_5(w)=\left(\frac{ (6w+H)^2}{9}-K \right)W^2=\left(\frac{ (6w+H)^2}{9}-K \right)\left(4 w^3+a w+b \right).
\]
Birational transformation destroys commutativity of  $z'_{1,2}$ and we have
\[
\{w_1,w_2\}\neq 0\,.
\]
Because  Abel's equations (\ref{w-ab-eq}) involve the reducible polynomial $\mathcal R_5(w)$, Kowalevski made an additional transformation
\bq\label{w-s}
w\to s+\frac{H}{3}
\eq
which reduce the standard Weierstrass equation  (\ref{ell-2}) for an elliptic curve  to  the following equation
\bq\label{ell-3}
E:\qquad S^2=4s^3 + 4Hs^2 + (4b^2c_1 + H^2 - K)s + 4b^2c_2^2\,,
\eq
which appears in the theory of degenerate Abel's integrals at $c_2\neq 0$.

The  Kowalevski variables of separation
\bq\label{kow-s}
s_{1,2}=w_{1,2}-\frac{H}{3}=\frac{R(z_1,z_2)\pm\sqrt{R(z_1,z_1)}\sqrt{R(z_2,z_2)}}{2(z_1-z_2)^2}\,,
\eq
satisfy Abel's equations
\bq\label{abel-kow}
\frac{\dot{s}_1}{\sqrt{\mathcal P_5(s_1)}}+\frac{\dot{s}_2}{\sqrt{\mathcal P_5(s_2)}}=0\,,\qquad \frac{s_1\dot{s}_1}{\sqrt{\mathcal P_5(s_1)}}+\frac{s_2\dot{s}_2}{\sqrt{\mathcal P_5(s_2)}}=1
\eq
on the hyperelliptic curve defining by the reducible polynomial $\mathcal P_5(s)$:
\bq\label{c-curve}
C:\qquad S^2=\mathcal P_5(s)\,,\qquad \mathcal P_5(s)= \Bigl((H+2s)^2 - K\bigr)S^2\,.
\eq
Variables $s_{1,2}$ are in the involution
\[\{s_1,s_2\}=0\]
 with respect to the Lie-Poisson brackets on the  Euclidean algebra $e^*(3)$. The corresponding canonically conjugated momenta, separation relations and $2\times 2$ Lax matrix with  the spectral curve $C$ (\ref{c-curve}) are discussed in \cite{kuz02}.

 Of course, Kowalevski never computed the Poisson brackets between variables $s_{1,2}$ and her reason for the transformation of variables (\ref{w-s})  was probably related to the investigations of degenerate Abel's integrals in \cite{kow84}, where she formulated so-called Weierstrass theorem on the periods of reducible integrals from his unpublished lectures \cite{coo84}.

  The meaning of the Kowalevski calculations has been discussed by many authors, see a shortlist  \cite{avm, fed16,kom90,kst03,kt05,kuz02,les88,mag18,wei83} and references within. Now any computer algebra system performs all these calculations for a few seconds.

\subsection{Abel's equations in $z'$-variables}
Birational transformation between  coordinates of the point $(z',Z')$ (\ref{z-var}) on $E$ (\ref{ell-1})  and coordinates $(s,S)$ of the same point on $E$ (\ref{ell-3})  reads as
\bq\label{s-z}
\begin{array}{rcl}
s& =&\dfrac{4b^2c_1 + 4bc_2z' + Hz'^2 - \sqrt{K-4b^2c_1\,}\,Z' - K}{2z'^2}\,,\\ \\
S&=&\dfrac{(2bc_2z'+4b^2c_1-K)Z' + \sqrt{K-4b^2c_1\,}(Hz'^2+ 6bc_2z' +4b^2c_1 - K)}{z'^3}\,.
\end{array}
\eq
Because
\[\{s_1,s_2\}=\{z'_1,z'_2\}=0\]
we have a canonical transformation of variables so that
${ds}/{S}={dz'}/{Z'}$ or
\[
\frac{ds}{\sqrt{4s^3 + 4Hs^2 + (4b^2c_1 + H^2 - K)s + 4b^2c_2^2}}=
\frac{dz'}{\sqrt{z'^4- 2Hz'^2 - 8bc_2z'  + K- 4b^2c_1}}\,.
\]
Thus, Abel's equations in $z'$-variables
\bq\label{abel-kow-z}
\frac{\dot{z}'_1}{\varphi(z'_1,Z'_1)} +\frac{\dot{z}'_2}{\varphi(z'_2,Z'_2)}=0\,,\qquad \frac{\dot{z}'_1}{\phi(z'_1,Z'_1)}+\frac{\dot{z}'_2}{\phi(z'_2,Z'_2)}=1
\eq
also have  two factors in the denominators:
\[
\varphi(z',Z')=Z'\sqrt{4s^2 +4Hs+H^2 - K\,}\,,\qquad \phi(z',Z')=Z's^{-1}\varphi(z',Z')\,,
\]
where $s$ is given by (\ref{s-z}).

We present here these simple calculations because variables of separation $z'_1$ and $z'_2$ do not satisfy the so-called Kowalevski conditions  \cite{mag15}. It allows us to say that these Kowalevski conditions are noninvariant conditions with respect to canonical transformations of variables preserving the additive separation of variables in the Hamilton-Jacobi equation.

\subsection{Degenerate Abel's integrals}
As early as 1832 Legendre had shown that  two hyperelliptic integrals
\[
\int\frac{dx}{\sqrt{x(1-x^2)(1-\varkappa^2x^2)}}\qquad \mbox{and}\qquad \int\frac{xdx}{\sqrt{x(1-x^2)(1-\varkappa^2x^2)}}
\]
are each expressible in terms of two elliptic integrals of the first kind through a quadratic transformation.
Immediately after,  Jacobi pointed out that this property holds for integrals on the hyperelliptic  curve
\[
\int\frac{dx}{\sqrt{R(x)}}\qquad \mbox{and}\qquad \int\frac{xdx}{\sqrt{R(x)}}
\]
defined by equation
 \bq\label{c-jac}
 C:\qquad y^2=R(x)\,,\qquad R(x)=x(1-x)(1-\kappa\lambda x)(1+\kappa x)(1+\lambda x)\,.
 \eq
 Invariant conditions for reducibility of abelian integrals involve the so-called Weierstrass-Picard theorem on the periods of reducible
integrals. This theorem was formulated by Picard \cite{pic83} in 1883 and  Kowalevski \cite{kow84} in 1884, see discussion in \cite{coo84}. Thus, it is clear that  Kowalevski was familiar with the Weierstrass theory of reducible Abel's integrals on hyperelliptic curves, which generalize the Jacobi calculations.

The Jacobi hyperelliptic curve $C$ (\ref{c-jac})  is isomorphic to a curve with an affine equation
    \[
C:\qquad   y^2=x^6-c_1x^4+c_2x^2-1\,,
  \]
having standard elliptic involutions $\sigma_{1,2}$, see detailed calculations in \cite{cass}.  The quotients $E_i =C/\sigma_i$ are two elliptic curves
 \begin{equation}\label{e-curve}
  E_1 :\quad  y^2 = x^3 - a_1x^2 + a_2x - 1\quad\mbox{and}\quad E_2 :\quad y^2 = x (x^3 - a_1x^2 + a_2x - 1)
  \end{equation}
so that Jacobian of $C$ decomposes  as $Jac(C)= E_1\times E_2$.   Now such elliptic fibrations of the reducible abelian varieties
  are studied extensively due to the promising post-quantum cryptography applications \cite{bes19}.

Let us come back to the Kowalevski top:
\begin{itemize}
  \item First curve $E_1$ in (\ref{e-curve}) coincides with the Kowalevski elliptic curve $E$ (\ref{ell-3}) at the special values of parameters $b$ and $c_2\neq 0$;
  \item Second curve $E_2$ in (\ref{e-curve}) is related to a spectral curve $\Gamma$ of the $4\times 4$ Lax matrix proposed by  Reyman and Semenov-Tian-Shansky  \cite{rs87} at  $c_2\neq 0$.
\end{itemize}
According to Theorem 7.8 in \cite{brs99}  if $c_2\neq 0$ the common level surface of the integrals of motion $H,K,c_1,c_2$ consists of two components (Liouville tori). If $c_2=0$ the level surface is irreducible.

To continue Kowalevski calculations in \cite{kow84,kow89}  we can consider two-dimensional abelian variety $E_1\times E_2$ and the corresponding group law on this reducible algebraic group.
For instance,  when an elliptic curve is realized as a nonsingular cubic curve, its group structure can be described in terms of the sets of three points in which lines intersect the curve. In our case
two points $P_1=(s_1,S_1)$ and $P_2=(s_2,S_2)$ determine a line
\bq\label{s-line}
\Upsilon:\qquad S=\frac{s - s_2}{s_1-s_2}S_1 +\frac{ s - s_1}{s_2-s_1}S_2\,,
\eq
 which has an intersection  with the elliptic curve $E$ (\ref{ell-3}) at the third point  $P_3=(s_3,S_3)$ so that
 \[
 \mbox{det} \left(
              \begin{array}{ccc}
                s_1 & S_1 & 1 \\
                s_2 & S_2 & 1 \\
                s_3 & S_3 & 1 \\
              \end{array}
            \right)=0\]
and
 \bq\label{ell-diff}
 \varphi_1(\dot{s}_1, \dot{s}_2, ,\dot{s}_3,c_1,c_2,H,K)= \frac{ds_1}{S_1}+\frac{ds_2}{S_2}+\frac{ds_3}{S_3}=0\,.
\eq
Affine coordinates of the third point $P_3$ are equal to
\[s_3=-(s_1+s_2)-H+\dfrac{(S_1-S_2)^2}{4(s_1-s_2)^2}=\dfrac{\eta}{\nu}\]
where
\[\begin{array}{rcl}
\eta&=& b^2\ell_2^2(\gamma_1-\mathrm i \gamma_2)^2 - 2\mathrm i(\ell_3z_2 + b\gamma_3)\bigl(b(\gamma_1-\mathrm \gamma_2) + z_2^2\bigr)\ell_2\ell_3 - (\ell_3^2 + z_2^2)(\ell_3z_2 + b\gamma_3)^2\,,\nn
\\
\\
\nu&=& \bigl(2b(\gamma_1-\mathrm i\gamma_2) + z_2^2\bigr)\ell_2^2 + (\ell_3 z_2+ b\gamma_3)^2\,,\nn
\end{array}
\]
and
\[S_3)=\frac{s_3 - s_2}{s_1-s_2}S_1 +\frac{ s_3 - s_1}{s_2-s_1}S_2\,.\]
Point $P_3=(s_3,S_3)$ is the desired reduced divisor $\rho(D')$ on the Kowalevski elliptic curve $E$ (\ref{ell-3}). Affine coordinates of this reduced divisor are dynamical variables which satisfy equations
\[\frac{ \{s_3,H\} }{2S_3}=\frac{ \ell_3z_2 - b\gamma_3 }{ \sqrt{\eta} }\qquad\mbox{and}\qquad
\frac{\{s_3,K\} }{4S_3}=\frac{ (2\ell_2\ell_3-\ell_3z_2 - b\gamma_3)\bigl(2b(\gamma_1-\mathrm i\gamma_2) + z_2^2\bigr)}{\sqrt{\eta} }\,.
\]
Using these equation we can easy check equation (\ref{ell-diff}) with elliptic differentials and obtain two  equations
\[
\varphi_2(\dot{s}_1, \dot{s}_2, ,\dot{s}_3,c_1,c_2,H,K)=0\qquad\mbox{and}\qquad \varphi_3(\dot{s}_1, \dot{s}_2, ,\dot{s}_3,c_1,c_2,H,K)=1
\]
similar to three equations for the three poles of the Backer-Akhiezer function associated with $4\times 4$ Lax matrix \cite{brs99}.

Only new thing is the construction of the $2\times 2$ Lax matrix on the Kowalevski elliptic curve $E=E_1$, which is a factor of reducible abelian variety $E_1\times E_2$.
Let us compute  Mumford's coordinates of the intersection divisor $D_3=P_1+P_2+P_3=(U,V)$ \cite{jac32d,mum84}
\bq\label{cub-kow-s}
U(s)=(s-s_1)(s-s_2)(s-s_3)\qquad\mbox{and}\qquad
V(s)=\frac{s - s_2}{s_1-s_2}S_1 +\frac{ s - s_1}{s_2-s_1}S_2+f(s)U(s)
\eq
where $f(s)$ is a rational function on $E$ without poles in $s_k$. Sometimes Mumford's coordinates of the divisor $D$ are called the Jacobi-Mumford polynomials themselves $(U, V)$, and sometimes Mumford's coordinates are called the coefficients of these polynomials $U$ and $V$. Both these polynomials $U$ and $V$ appeared in Abel's memories \cite{ab}, when he also used rational functions instead of polynomials.

Using these polynomials we can construct a new $2\times 2$ Lax matrix for the Kowalevski top
\[
L(s)=\left(
       \begin{array}{cc}
         V & U \\
         W & -V\\
       \end{array}
     \right)\,,\qquad W=\frac{S^2-V^2}{U}\,,
\]
where $S$ is given by (\ref{ell-3}).  Thus, using reduced divisor we can get a new formal Lax matrix on the first factor $E_1=E$ of the reduced abelian variety  $E_1\times E_2$ \cite{ts21}. The corresponding Backer-Akhiezer function  $\psi$ (\ref{ba-f}) with  the standard normalization $\vec{\alpha}$ has three poles $P_1$, $P_2$ and $P_3$, which lie on the line $\Upsilon$ (\ref{s-line}).
Spectral curves of the similar $2\times 2$ Lax matrices from \cite{per} and \cite{kuz02} coincide with the Kowalevski hyperelliptic curve $C$ (\ref{c-curve}).

Of course, variables $s_{1,2}$ and the corresponding Abel's equations (\ref{abel-kow})  are well studied. Nevertheless, in the  literature, we do not find a discussion of the  group operations on a reducible abelian variety $E_1\times E_2$ generating the following transformations of the Kowalevski variables
\[
\tau:\qquad (s_1,S_1)+(s_2,S_2)\to (s'_1,S'_1)+(s'_2,S'_2)
\]
 so that
 \begin{enumerate}
  \item $(s'_1,S'_1)$ and $(s'_2,S'_2)$ are points on the first factor $E_1$;
  \item $(s'_1,S'_1)$ and $(s'_2,S'_2)$ are points on the second  factor $E_2$;
  \item $(s'_1,S'_1)$ and $(s'_2,S'_2)$ are points on $E_1$ and $E_2$, respectively.
\end{enumerate}
The group law on the reducible abelian variety $E_1\times E_2$ is beyond the scope of this note dedicated to the reduced divisors on elliptic curves $E_1$ and $E_2$.

\section{Reduced divisor on the spectral curve of Lax matrix}
\setcounter{equation}{0}
In \cite{rs87}  Reyman and  Semenov-Tian-Shansky found Lax matrices for the  Kowalevski top. In   \cite{brs99} these  Lax matrices were used to integrate the problem in terms of theta-functions, see also textbook \cite{rs}.

Let us take Lax matrix $L$  (6.3) from  \cite{brs99} and multiply it's first term on $b$ that corresponds to scaling $g_i\to bg_i$. As a result, we obtain the Lax matrix
\begin{eqnarray}\label{lax}
L(\lambda)=\frac{\mbox{\rm i}b}{\lambda}\left(
                                    \begin{array}{cccc}
                                      0 & g_1-\mbox{\rm i}g_2 & 0 & -g_3\\
                                      -g_1-\mbox{\rm i}g_2 & 0 & g_3 & 0 \\
                                      0 & -g_3 & 0 & -g_2-\mbox{\rm i}g_2 \\
                                      g_3 & 0 & g_1-\mbox{\rm i}g_2 & 0 \\
                                    \end{array}
                                  \right) \\
\nonumber\\
+\mbox{\rm i}\left(
                 \begin{array}{cccc}
                   0 &0 & \ell_1-\mbox{\rm i}\ell_2 &0 \\
                   0 & 0 & 0 & -\ell_1-\mbox{\rm i}\ell_2 \\
                   \ell_1+\mbox{\rm i}\ell_2 & 0 & -2\ell_3 & -2\lambda \\
                   0 &-\ell_1+\mbox{\rm i}\ell_2& 2\lambda & 2\ell_3 \\
                 \end{array}
               \right)\,,\quad \mbox{\rm i}=\sqrt{-1}\,.
 \nonumber
\end{eqnarray}
with the spectral curve  defined by equation
\bq\label{rs-curve}
\Gamma:\qquad \mbox{det}\left( L(\lambda)-\mu\right)=\mu^4-2\left(2\lambda^2 -H +\dfrac{b^2c_1}{\lambda^2}\right)\mu^2 + K
 -\dfrac{2b^2\left(c_1H- 2c_2^2\right)}{\lambda^2} + \dfrac{b^4c_1^2}{\lambda^4}=0\,,
\eq
where $c_{1,2}$ are the Casimir functions (\ref{caz-fun}) and  integrals of motion $H$ and $K$ are given by (\ref{h}-\ref{k}).

Symmetries of Lax matrices give rise to the two commuting involutions $\tau_1$ and $\tau_2$ on $\Gamma$, that allows us to consider quotient elliptic curve $\mathcal E =\Gamma/(\tau_1,\tau_2)$ \cite{brs99}. Indeed, substituting
\[
\mu^2=v\,,\qquad\mbox{and}\qquad \lambda^2=u
\]
into (\ref{rs-curve})  we obtain the following equation
 \bq\label{ell-4}
\mathcal  E:\qquad
\Phi(u,v)= (uv)^2 -2\left(2u^2+Hu+b^2c_1\right)uv + Ku^2 - 2b^2\left(c_1H - 2c_2^2\right)u + b^4c_1^2=0\,,
 \eq
 which after  birational transformation
 \bq\label{bi-rat-k}
 v\to \dfrac{y-2u^2-Hu-b^2c_1}{u}\qquad\mbox{and}\qquad u=-x
 \eq
looks like
 \bq\label{ell-5}
\mathcal E:\qquad y^2=xS^2\,, \qquad S^2=4x^3 + 4Hx^2 + (4b^2c_1 + H^2 - K)x + 4b^2c_2^2\,.
\eq
Here $S^2$ is the cubic polynomial from the definition of the  Kowalevski elliptic curve $E=E_1$   (\ref{ell-3}).

At $c_2\neq 0 $ we can identify curve  $\mathcal E$ with  the second factor $E_2$ (\ref{e-curve})  in the reduced abelian variety $E_1\times E_2$.

\subsection{Standard normalization and $r$-matrix}
Let us  impose a linear constraint
\[
\vec{\alpha} \cdot \psi=\sum_{i=1}^N \alpha_i\psi_i=1\,,
\]
i.e. fix  normalization $\vec\alpha$ of the Backer-Akhiezer function $\psi $ (\ref{ba-f}).  Poles of  $\psi$ form a divisor $D$ on the spectral curve $\Gamma$ (\ref{s-curve}).
Following the Sklyanin scheme, we substitute matrices
\[
L^{(p)}= L\left({\rm tr}\,L^{(p-1)}\right) - (p-1)\, L^{(p-1)} L\,,\quad\mbox{with}\quad L^{(1)}\equiv L,
\]
labelled by number $p=1,\cdots,N$ into the $N\times N$ matrix
\bq\label{b-mat}
{\cal B}(x)=
\left(\begin{array}{c}
        \vec\alpha\cdot L^{(1)}(x)\;L^{-1}(x)\\
         \vec\alpha\cdot L^{(2)}(x)\;L^{-1}(x)\\
      \dfrac12\;\vec\alpha\cdot L^{(3)}(x)\;L^{-1}(x)\\
       \cdots\\
      \dfrac{1}{(N-1)!}\;\vec\alpha\cdot L^{(N)}(x)\;L^{-1}(x)\\
\end{array}\right)\,,
\eq
so that
\[
\vec\alpha\cdot(L(x)-y)^\wedge
\equiv\left((-y)^{N-1},(-y)^{N-2},\ldots,1\right)\cdot\,{\cal B}(x)=0\,.
\]
This matrix determine first  Mumford's coordinate of  the divisor $D=(U(x), V(x))$
\bq\label{U-coord}
U(x)=\mbox{MakeMonic}\,\det {\cal B}(x)\,,
\eq
and  a finite set of second Mumford's coordinates
\[y^m-V_m(x)=0\,,\qquad y_i^m=V_m(x_i)\,,\qquad i=1,\ldots,N\,,\qquad m=1,\ldots,N-1\,,\]
see \cite{kuz97,kuz022,skl95} and references within. As usual, second Mumford's coordinates
\bq\label{V-coord}
(-y)^{j-i}=\frac{({\cal B}^\wedge(x))_{k,i}}{({\cal B}^\wedge(x))_{k,j}}\,,
\eq
 are equivalent up to mod $U(x)$, see (\ref{cub-kow-s}), and the MakeMonic means that we take the numerator of a rational function and divide the corresponding polynomial by its leading coefficient.

For the $4\times 4$ Lax matrix  $L$ (\ref{lax}) and  the standard normalization vector
\bq\label{st-norm}
\vec\alpha=(1,0,0,0)\,,
\eq
 matrix $\cal B$ (\ref{b-mat}) is equal to
\[
 {\cal B}(\lambda)=\left(
               \begin{array}{cccc}
                \scriptstyle  1 & \scriptstyle 0 &  \scriptstyle 0 &  \scriptstyle 0 \\
                \scriptstyle  0&\frac{- \mbox{\rm i}bg_-}{\lambda}& {\scriptstyle -\mbox{\rm i}\ell_-}&\frac{\mbox{\rm i}bg_3}{\lambda}\\
               {\scriptstyle  -4\lambda^2 + \ell_1^2 + \ell_2^2 + 4\ell_3^2 - 4bg_1 -}\frac{b^2c_1}{\lambda^2}&\scriptstyle 0& {\scriptstyle 2\ell_3\ell_- + 2bg_3}& {\scriptstyle 2\ell_-\lambda +} \frac{2bc_2}{\lambda}\\
               {\scriptstyle   -2i(\ell_1^2\ell_3 + 2bg_3\ell_1 + \ell_2^2\ell_3) -} \frac{2\mbox{\rm i}b^2g_3c_2}{\lambda^2} &\scriptstyle b_{24}  &\scriptstyle b_{34}  &\frac{2\mbox{\rm i}b^2g_3g_-+\mbox{\rm i}b\ell_+\bigl(g_3\ell_- - 2g_-\ell_3 \bigr)}{\lambda}
-\frac{\mbox{\rm i}b^3g_3c_1}{\lambda^3}  \\
               \end{array}
             \right)
\]
where
\[\begin{array}{rcl}
b_{24}&=& 2\mbox{\rm i}\left(\ell_-^2 +2bg_-\right)\lambda
+\dfrac{b\left(
4\mbox{\rm i}\ell_3(g_3\ell_ --g_-\ell_3) + 2\mbox{\rm i}b (g_3^2 +2g_1g_-) +\mbox{\rm i}g_+\ell_-^2\right)
}{\lambda}
+\dfrac{\mbox{\rm i}b^3c_1g_-}{\lambda^3}\,,\\
\\
b_{34}&=&-\mbox{\rm i}\ell_+\left(\ell_-^2+2bg_- \right)+
\dfrac{\mbox{\rm i}b^2\left(
 g_3^2\ell_--2g_-g_3\ell_3 -g_-^2\ell_+\right)}{\lambda^2}\,,
\end{array}
\]
and
\[
g_\pm=g_1\pm \mbox{\rm i}g_2\,,\qquad \ell_\pm=\ell_1\pm \mbox{\rm i}\ell_2\,,
\]
The corresponding first Mumford's coordinate $U(\lambda)$ (\ref{U-coord})  has the following form
\bq\label{b3-kow}
U(\lambda)=\lambda^6+\mathrm{u}_4\lambda^4+\mathrm{u}_2\lambda^2+c_2\mathrm{u}_0\,,
\eq
where coefficients  $\mathrm{u}_k$ may be recovered from the definition  (\ref{U-coord}) or found in \cite{kuz022}.

 In 2002 Kuznetsov  proved  that  roots  $\lambda_{1,2}$ of the polynomial $U(\lambda) $ (\ref{b3-kow}) at $c_2=(\ell.\gamma)=0$
\[U(\lambda)=\lambda^2(\lambda^2-\lambda_1^2)(\lambda^2-\lambda_2^2)\]
are variables of separation in the corresponding  Hamilton-Jacobi equation    \cite{kuz022}. Thus,  poles of the Backer-Akhiezer function  with the  standard normalization  (\ref{st-norm}) are
the variables of separation obtained in the framework of the Sklyanin method.

In 2018 Dubrovin and  Skrypnyk introduced the necessary and sufficient conditions for satisfiability of the Sklyanin scheme in the terms of the classical $r$-matrix  \cite{dub18}.  It is easy to show that $r$-matrix for the Kowalevski top \cite{rs87} does not satisfy these conditions at any value of $c_2$. Thus, according to \cite{dub18} poles of the Backer-Akhiezer function with numerical normalization $\vec{\alpha}$ can not generate variables of separation in the framework of the Sklyanin method. Nevertheless,  these variables  exist and we can say that the Dubrovin-Skrypnyk theory is wrong at least in the Kowalevski case.

At $c_2=0$ we have irreducible abelian variety and it is a principal cause for applicability of the Sklyanin scheme which is independent on the $r$-matrix describing the Poisson brackets between entries of the Lax matrix. The mapping of the Liouville torus to Jac$C$ becomes an unramified two-sheeted covering, see Sect. 7.9 in \cite{brs99}.

Another variables of separation in the partial case at $c_2=0$ were proposed in \cite{ts02,ts10}.   Relations between various algebraic curves for the Kowalevski top are discussed in \cite{avm,fed16, en18,mark01}.

\subsection{Semi-reduced  divisors}
Below we study the only  case
\[c_2=\ell_1g_2+\ell_2g_2+\ell_3g_3\neq 0\]
when we can construct reducible abelian variety $E_1\times E_2$. In this case evolution of the three poles $P_1,P_2$ and $P_3$ of the Backer-Akhiezer function
 is defined by three equations involving elliptic and Prym differentials, see equations (7.69)  in \cite{brs99}. An interesting problem is how to
 compute the action-angle variables of the Kowalewski top \cite{du98} directly from  these three poles of the Backer-Akhiezer function \cite{brs99}.

At $c_2\neq 0$ and $\lambda^2=u$ we have a cubic polynomial $U(u)$ (\ref{b3-kow})  or  first Mumford's coordinate of the  divisor of poles on the elliptic curve $\mathcal E$
\[D=P_1+P_2+P_3\,.\]
According to  the Riemann-Roch theorem dimension of the  linear system $|D|$, which  is the set of all nonnegative divisors which are linearly equivalent to $D$
 \[
|D|=\{D'\in \mathrm{Div}(C)\,|\,D'\sim D\,\mathrm{and}\,D'>0\}\,,
\]
is equal to
\[\mathrm{dim} |D|=\mbox{deg}\,D-g\,,\quad\mbox{at}\quad \mbox{deg}D=n>g\,,\]
where $g$ is a genus of $C$ and deg$D$ is a degree of the divisor, see definitions and other details in textbook \cite{mir95}.
In our case
\[\mathrm{dim} |D|=\mbox{deg}\,D-g(\mathcal E)=3-1=2\,,\]
and we have nontrivial space of equivalent divisors involving a chain of divisors
\[
D\to D'\to D''\,, \qquad \mbox{deg}D'=2\,,\quad \mbox{deg}D''=1\,.
\]
We aim to construct these divisors and to study the evolution of these divisors.

Let us calculate six possible second Mumford's coordinates (\ref{V-coord}) for $D=(U(x),V(x))$
\[
\mu^2-V_2(\lambda)\,,\qquad V_2(\lambda)=\frac{\left({\cal B}^\wedge(\lambda)\right)_{k,j}}{({\cal B}^\wedge(\lambda))_{k,j+2}}\,,\qquad j=1,2,\qquad k=2,3,4.
\]
Substituting $\mu^2=V_2(\lambda)$ into the equation for spectral curve $\Gamma$ (\ref{rs-curve})  we obtain  rational function on $\lambda$
\[
V_2^4(\lambda)-2d_1(\lambda)V_2(\lambda)+d_2(\lambda)=0\,.
\]
Its numerator is so-called Abel's polynomial \cite{ab,jac32d}
\[
\Psi(\lambda)=\theta\, U(\lambda)U'(\lambda)=0\,,
\]
generating  coordinates of divisors $D$ and $D'$ so that
\[
D+D'+D_\infty=0\,,
\]
where $D_\infty$ is a linear combination of the points at infinity. It is easy to prove that
\begin{itemize}
  \item at $k=3,4$ and $j=1$ divisor $D'$ has degree more then degree of divisor $D$;
  \item at $k=2,i=1$ and  $k=3,j=2$ or  $k=4,j=2$  divisor $D'$ is a constant divisor of degree two with coordinate
\[
U'(\lambda)=\lambda^2;
\]
  \item at $k=j=2$ divisor $D'=(U'(x),V'(x)$ has degree four and its first Mumford's coordinates is equal to
\[
U'(\lambda)=\lambda^4-\left(\ell_3^2 + \frac{2b\ell_1(g_1\ell_1 +g_2\ell_2)+b^2g_3^2}{\ell_1^2+\ell_2^2}\right)\lambda^2
+\frac{b^2c_2^2}{\ell_1^2+\ell_2^2}\,.
\]
\end{itemize}
Entries of the corresponding normalization $\vec\alpha'$ are bulky functions on $e^*(3)$ that we will omit for brevity.

Summing up, using numerical normalization vector $\vec\alpha$   and Abel's reduction of divisors we  obtain divisor of degree two $D'=(U'(x), V'(x)$ on the elliptic curve $\mathcal E$ (\ref{ell-4})  with following Mumford's coordinates
\bq\label{u-coord}
U'(u)=(u-u_1)(u-u_2)\equiv u^2-\left(\ell_3^2 + \frac{2b\ell_1(g_1\ell_1 +g_2\ell_2)+b^2g_3^2}{\ell_1^2+\ell_2^2}\right)u
+\frac{b^2c_2^2}{\ell_1^2+\ell_2^2}\,.
\eq
and
\bq\label{v-coord}
v-V'(u)\,,\qquad V'(u)=\left.\frac{({\cal B}^\wedge(\lambda))_{2,2}}{({\cal B}^\wedge(\lambda))_{2,4}}\right|_{\lambda^2=u}\,.
\eq
Because
\[
\mbox{dim}|D'|=\mbox{deg}D'-g(\mathcal E)=2-1=1\,,
\]
divisor $D'$ is a semi-reduced divisor on the elliptic curve, which can be reduced to the one-degree equivalent divisor $D''$.

\subsection{Reduced divisor}
Birational transformation $(v,u)\to (x,y)$ (\ref{bi-rat-k}) transforms elliptic curve $\mathcal E$ to canonical form (\ref{ell-5}) that allows as to directly apply  Abel's  calculations \cite{ab}. Indeed, two points $P_1'=(x_1,y_1)$ and $P_2'=(x_2,y_2)$ of divisor $D'$ on an elliptic curve
\[\mathcal E:\qquad y^2=a_4x^4+a_3x^3+a_2x^2+a_1x+a_0\]
determine parabola
\[
\Upsilon':\qquad
y=\sqrt{a_4}x^2+b_1x+b_0=\sqrt{a_4}(x-x_1)(x-x_2)+\dfrac{(x-x_2)y_1}{x_1-x_2}+\dfrac{(x-x_1)y_2}{x_2-x_1}\,,
\]
which has a finite third point of intersection $P'_3=(x_3,y_3)$ with affine coordinates
\bq\label{x3}
x_3= -x_1 - x_2 -\frac{ 2\sqrt{a_4} b_0+ b_1^2 - a_2}{2\sqrt{a_4}b_1 - a_3}\,,\qquad y_3=\sqrt{a_4}x_3^2+b_1x_3+b_0\,.
\eq
These equations describe the reduction of divisors
\[
D'=P_1'+P_2'\to D''=P_3'\,.
\]
In our case, a unique reduced divisor $D''$ in a class of equivalent divisors associated with standard normalization $\vec\alpha$ (\ref{st-norm}) consists of the point
\[
\rho(D)=D''=P'(x_3,y_3)\,,\qquad \mbox{dim}|D''|=\mbox{deg}D''-g(\mathcal E)=1-1=0\,,
\]
with affine coordinates
\[
x_3=\frac{4b^2c_2^2}{4b^2c_1 - K}\,,\qquad y_3=\frac{4b^2c_2^2(4b^2c_1H - 8b^2c_2^2 - HK)}{(4b^2c_1 - K)^2}\,.
\]
Entries of the corresponding normalization $\vec\alpha''$ are bulky functions on $e^*(3)$ that we  will omit for brevity.

Affine coordinates of the reduced divisor $\rho(D)$ can be considered as action coordinates on the phase space $e^*(3)$ which were obtained from the three poles $P_1,P_2$ and $P_3$ of the Backer-Akhiezer function by a standard reduction procedure.

\subsection{Evolution of semi-reduced divisor}
Two variable  points $P_1'$, $P_2'$ and one fixed point $P'_3$  lie on  the parabola $\Upsilon'$ and, therefore, so-called geometric Abel's integral \cite{ab}
\[
\frac{dx_1}{y_1}+\frac{dx_2}{y_2}+\frac{dx_3}{y_3}=0\,,
\]
determines evolution of the points $P_1'(x_1,y_1)$ and $P'_2(x_2,y_2)$ on elliptic curve
\bq\label{g-eq1}
\frac{dx_1}{y_1}+\frac{dx_2}{y_2}=0\,.
\eq
So, similar to the Kepler problem and harmonic oscillator \cite{ts19s,ts20r}, we have an evolution of the parabola $\Upsilon'$  around a fixed point on an elliptic curve.

Abscissas $u_{1,2}$ (\ref{u-coord}) and ordinates (\ref{v-coord})
\[
v_{1,2}=V'(u=u_{1,2})\,,
\]
 of the points in  support of the semi-reduced divisor $D'$ satisfy to equation  (\ref{ell-4}) for the elliptic  curve $\mathcal E\cong E_2$
\[\Phi(u_i,v_i)=0\,.\]
 Variables $u_{1,2}$ do not commute to each other
\[\{u_1,u_2\}\neq 0\,,\]
similar to $w_{1,2}$ variables in the first component $E=E_1$ of the reducible abelian variety $E_1\times E_2$.

Affine coordinates of poles  $u_{1,2}$ and  $v_{1,2}$ satisfy to the following differential equations
\bq\label{eq-uv}
\begin{array}{r}
\Omega_1(u_1,v_1)\dot{u}_1+\Omega_1(u_2,v_2)\dot{u}_2=0\,,\\
\\
 \Omega_2(u_1,v_1)\dot{u}_1+\Omega_2(u_2,v_2)\dot{u}_2=0\,,
 \end{array}
\eq
on the elliptic curve $\mathcal E$ (\ref{ell-4}). Here
\[
\dot{u}_k=\{H,u_k\}
\]
and
\bq\label{eq-uv-k}
\begin{array}{rcl}
\Omega_1(u,v)&=&\dfrac{1}{u}\,\dfrac{\partial_H \Phi(u,v)}{\partial_u \Phi(u,v)}=\dfrac{b^2c_1 - uv}{u(b^2c_1 - Hu - uv + 2u^2)}\,,
\\
\\
\Omega_2(u,v)&=&\dfrac{1}{u}\,\dfrac{\partial_K \Phi(u,v)}{\partial_u \Phi(u,v)}=-\dfrac{1}{2(b^2c_1 - Hu - uv + 2u^2)}\,.
\end{array}
\eq
Birational transformation $(v,u)\to (x,y)$ (\ref{bi-rat-k}) transforms elliptic curve $\mathcal E$ to canonical form (\ref{ell-5})
\[y^2=xP_3(x)\,,\qquad P_3(x)=4x^3 + 4Hx^2 + (4b^2c_1 + H^2 - K)x + 4b^2c_2^2\]
and also reduces equations (\ref{eq-uv}) to  Abel's equations
\bq
\label{eq-xy}
\begin{array}{c}
\dfrac{dx_1}{y_1}+\dfrac{dx_2}{y_1}=0\,,\\ \\
\left(\dfrac{dx_1}{x_1}+\dfrac{2x_1dx_1}{y_1}\right)+\left(\dfrac{dx_2}{x_2}+\dfrac{2x_2dx_2}{y_2}\right)=0\,,
\end{array}
\eq
involving holomorphic 1-form $\omega_1$  and  logarithmic 1-form $\omega_2$
\[\omega_1=\frac{dx}{y}\,,\qquad \omega_2=\frac{dx}{x}-\frac{2xdx}{y}\,.\]
A differential of the third kind $\omega_3=1/xy$ appears in the following  equation
\bq\label{eq-xy-3}
\dfrac{\dot{x}_1}{x_1y_1}+\dfrac{\dot{x}_2}{x_2y_2}=\dfrac{1}{2}\dfrac{d}{dt} \dfrac{1}{x_1x_2}\,.
\eq
Equation involving time involves Prym differentials on the corresponding double covering of elliptic curve $\mathcal E=E_2$.

It is easy to see that independent on the phase space variables $x_{1,2}$ become dependent on the common level of integrals of motion because
\[
(x+x_1)(x+x_2)=x^2 + \frac{X(X-2H)+K-4c_1b^2 }{4X}\,x + \frac{c_2^2b^2}{X}\,,\qquad X=z_1z_2\equiv\ell_1^2+\ell_2^2\,.
\]
Thus,  equations (\ref{eq-xy}) and (\ref{eq-xy-3}) on a second factor $\mathcal E=E_2$ of a reducible abelian variety $E_1\times E_2$
 describe the evolution of the one variable $X$  similar to the Clebsch case \cite{ts21a}.

\subsection{Kowalevski gyrostat}
Our calculations can be directly generalized to the Kowalevski gyrostat with the Lax matrix
\[
\tilde{L}=L+\mbox{\rm i}\gamma\,\mathrm{diag}(-1, 1,-1,1)\,,\qquad \gamma\in \mathbb R\,,
\]
and the following integrals of motion
\[\begin{array}{rcl}
\tilde{H}&=& \ell_1^2 + \ell_2^2 + \ell_3^2 + (\ell_3 + \gamma)^2 - 2bg_1\,,\\
\\
\tilde{K}&=& (4g_1^2 + 4g_2^2)b^2 + (4\ell_1^2g_1 + 8\ell_1\ell_2g_2 - 8\gamma\ell_1 g_3 - 4\ell_2^2g_1 - 4\gamma^2g_1)b + (\ell_1^2 + \ell_2^2 - 2\ell_3\gamma - \gamma^2)^2\,.
\end{array}
\]
The corresponding spectral curve $\tilde{\Gamma}$ is also twofold coverings of the elliptic curve $\tilde{\mathcal E}$ defined by the equation
\[
 \tilde{\Phi}(u,v)=\Phi(u,v)-4\gamma^2 u^3=0\,,
\]
whereas first  Mumford's coordinate  for semi-reduced divisor $\tilde{D}'$ on $\tilde{\mathcal E}$ is equal to
\[
\tilde{U}'(u)=(u-\tilde{u}_1)(u-\tilde{u}_2)=U'(u)-\gamma u\left(2\ell_3+\gamma+\dfrac{2b\ell_1 g_3}{\ell_1^2 + \ell_2^2}\right)\,.
\]
As above coordinates of poles  $\tilde{u}_{1,2}$ and $\tilde{v}_{1,2}$ satisfy equations of the form (\ref{eq-uv}) and (\ref{eq-uv-k}) on the elliptic curve $\tilde{\mathcal E}$.  Similar equations appear also for other generalizations of the Kowalevski curve associated with elliptic curves \cite{brs99,kom90,kst03}.

\section{Conclusion}
 We  construct  and study reduced divisors  associated with the Kowalevski variables of separation and the poles
 of the  Baker-Akhiezer function with a  standard normalization.

 Affine coordinates of the first reduced divisor are dynamical variables which can be useful to a description of the group law on the reducible abelian variety $E_1\times E_2$, where elliptic curves  $E_1$ and $E_2$ were introduced by  Kowalevski \cite{kow89}  and by Reyman and Semenov-Tian-Shansky up to isogenies \cite{rs87}, respectively. For instance, we can use this reduced divisor  to construct
 $2\times 2$ Lax matrix on the first factor in the reducible abelian variety $E_1\times E_2$, i.e. spectral curve of this Lax matrix is the Kowalevski elliptic curve.

Classical $r$-matrix for the  $4\times 4$ Lax matrix introduced by  Reyman and Semenov-Tian-Shansky does not satisfy to the Dubrovin-Skrypnyk necessary and sufficient conditions for satisfiability of the Sklyanin construction of the variables of separation in the Hamilton-Jacobi equation. Nevertheless, Kuznetsov proved that poles of the Baker-Akhiezer function with a standard normalization are variables of separation when one of the Casimir functions on $e^*(3)$ is equal to zero. Thus, the Dubrovin-Skrypnyk theory is wrong, at least in the Kowalevski case.

 It was one of the reasons to study reduced representative of a class of equivalent divisors on an elliptic curve determined by poles of the  Baker-Akhiezer function with a  standard normalization.
 If one of the Casimir functions is equal to zero, the spectral curve of the $4\times 4$ Lax matrix determines irreducible abelian variety and the Sklyanin scheme works. If this Casimir function is nonzero we can consider the reducible abelian variety and the corresponding reduced divisor on the second elliptic curve in a product $E_1\times E_2$. Affine coordinates of this reduced divisor can be considered as action variables, which are variables of separation with trivial dynamics.

As a by-product, we obtain equations describing the evolution of the semi-reduced divisor of degree 2 on the elliptic curve $E_2$ defined by a spectral curve of the $4\times 4$ Lax matrix. We plan to discuss these equations on the elliptic curve $E_2$ and similar equations on the Kowalevski elliptic curve $E_1$ in the forthcoming publication.


\begin{thebibliography}{10}
\bibitem{ab}
 Abel N. H.,
 \newblock{\em M\'{e}moire sure une propri\'{e}t\'{e}
g\'{e}n\'{e}rale d'une classe tr\`{e}s \'{e}ntendue de fonctions transcendantes},
Oeuvres compl\'{e}tes, Tome I, Grondahl Son, Christiania,  pp. 145-211, (1881).

\bibitem{avm}
 Adler M, van Moerbeke P.,
 \newblock{\em  The Kowalewski and Henon-Heiles motions as Manakov geodesic flows on SO(4) a two-dimensional family of Lax pairs}, Comm. math.Phys., v.113, pp. 659-700, (1988).

\bibitem{bes19}
Beshaj L., Elezi A., Shaska T.,
\newblock{\em Isogenous components of Jacobian surfaces}, European Journal of Mathematics, v.6, pp.1276-1302, (2020).

\bibitem{brs99}
Bobenko A.~I., Reyman A.~G., Semenov-Tian-Shansky M.~A.,
\newblock {\em The {K}owalewski top 99 years later: a {L}ax pair, generalizations and explicit solutions},
\newblock Comm. Math. Phys.,~v.122:2, pp.321-354, (1989).

\bibitem{cass}
Cassels J., Flynn V., Prolegomena to a Middlebrow Arithmetic of Curves
of Genus 2, London Mathematical Society Lecture Note Series, v.230, (1996).


\bibitem{coo84}
Cooke R, Degenerate Abelian Integrals. In: The Mathematics of Sonya Kovalevskaya. Springer, New York, NY, (1984).

\bibitem{du98}
Dullin H. R., Richter P. H., Veselov  A. P.,
\newblock{\em Action variables of the Kovalevskaya top}, Regul. Chaotic Dyn., v.3, pp. 18-31, 1998.

\bibitem{dub18}
Dubrovin B.,  Skrypnyk T.,
\newblock{\em Separation of variables for linear Lax algebras and classical r-matrices},   J. Math. Phys., v, 59,  091405, (2018).

\bibitem{en18}
Enolski V.Z., Fedorov Yu.N.,
 \newblock{\em Algebraic description of Jacobians isogenous to certain Prym varieties with polarization (1,2)},
 Experimental Mathematics, v.27, pp.147-178, (2018).


\bibitem{fed16}
Fedorov Yu.N., Garc\'{\i}a-Naranjo L.C., Joan C. Naranjo J.C.,
\newblock{\em A shortcut to the Kovalevskaya curve}, preprint  	arXiv:1606.08331, 2016.

\bibitem{jac32d}
Jacobi C. G. J.,
 \newblock{\em \"{U}ber eine neue Methode zur Integration der hyperelliptischen Diff\-eren\-tial\-glei\-chun\-gen und \"{u}ber die rationale Formihrer vollst\"{a}ndigen algebraischen Integralgleichungen}, J. Reine Angew. Math., v.{32}, pp.220-227, 1846.

\bibitem{kom90}
Komarov I.V.,  Kuznetsov V.B.,
\newblock{\em   Kowalewski’s top on the Lie algebras o(4), e(3) and o(3, 1)},  J. Phys. A, v.23, pp. 841-846,  (1990).

\bibitem{kst03} Komarov I.V.,  Sokolov V.V, Tsiganov A.V.,
 \newblock{\em Poisson maps and integrable deformations of Kowalevski top},
 \newblock{J. Phys. A.}, v.{36}, pp.8035-8048, (2003).

 \bibitem{kt05} Komarov I.V.,   Tsiganov A.V.,
 \newblock{\em  On a trajectory isomorphism of the Kowalevski gyrostat and the Clebsch problem},
 J. Phys. A: Math. Gen., v.38, pp.2917, (2005).

\bibitem{kow84}
 Kowalevski S.,
\newblock{\em
\"{U}ber die Reduction einer bestimmten Klasse Abel’scher Integrale 3ten Ranges auf elliptische Integrale},
\newblock{Acta Math}, v.4,  pp. 393-414, (1884).

\bibitem{kow89}
 Kowalevski S.,
\newblock{\em Sur le probl\'{e}me de la rotation d'un corps solide
autour d'un point fixe},
\newblock{Acta Math.}, v.{12}, pp.177-232, (1889).

\bibitem{kuz97}
 Kuznetsov V. B., Nijhoff  F. W.,   Sklyanin E. K.,
\newblock{\em  Separation of variables for the Ruijsenaars system},
Commun. Math. Phys., v.189, pp. 855-877, (1997).

\bibitem{kuz02}
Kuznetsov, V.B.,
\newblock{\em  Kowalevski top revisted}, CRM Proc. Lecture Notes 32, Providence, RI: Amer. Math. Soc.,  pp. 181-196, (2002).

\bibitem{kuz022}
 Kuznetsov V. B.,
 \newblock{\em Simultaneous separation for the Kowalevski and Goryachev-Chaplygin gyrostats},
  J. Phys. A: Math. Gen., v.35,  pp. 6419-6430, (2002).


 \bibitem{les88}
Lesfari, A.
\newblock{\em Abelian surfaces and Kowalewski's top},
Annales scientifiques de l'\'{E}cole Normale Sup\'{e}rieure, S\'{e}rie 4, Tome 21, no. 2, pp. 193-223, (1988).

  \bibitem{mag15}
Magri F., Skrypnyk T.,
 \newblock{\em The Clebsch system}, preprint, arXive:1512.04872, (2015).

\bibitem{mag18}
Magri F.,
\newblock{\em The Kowalevski top revised}, In {\em Integrable Systems and Algebraic Geometry}, v.1, pp.329-355, edited by Ron Donagi and  Tony Shaska, Cambridge University Press, (2020).

\bibitem{mark01}
Markushevich D., \newblock{\em Kowalevski top and genus-2 curves},
J. Phys. : Math. Gen., v. 34, pp. 2125-2135, (2001).

  \bibitem{mir95}
Miranda R.,
\newblock{Algebraic Curves and Riemann Surfaces}, Graduate Studies in Mathematics, vol 5., American Mathematical Society; UK ed., (1995).

\bibitem{mum84}
 Mumford D., Tata Lectures on Theta II, Birkh\"{a}user, (1984).

\bibitem{pa99}
Paulus, S., and R\"{u}ck, H.-G.,
\newblock{\em Real and imaginary quadratic representations of hyperelliptic function fields},
Math. Comp. v.68, n.227, pp.1233-1241, (1999).

\bibitem{per}
Perelomov A.M.,
\newblock{\em A Lax representation for systems of S. Kovalevskaya type},
 Funct. Anal. Appl., v.16:2, pp.143-144, (1982).


 \bibitem{pic83}
Picard E.,
\newblock{\em
 Sur la r\'{e}duction du nombre des p\'{e}riodes des int\'{e}grales ab\'{e}liennes et, en particulier, dans le cas des courbes du second genre
},
Bulletin de la Soci\'{e}t\'{e} Math\'{e}matique de France, v.11,  pp. 25-53, (1883).

\bibitem{poi84}
Poincar\'{e} H.,
\newblock{\em Sur la r\'{e}duction des int\'{e}grales ab\'{e}liennes},
Bulletin de la Soci\'{e}t\'{e} Math\'{e}matique de France, v.12, pp.124-143, (1884).

\bibitem{rs87}
Reyman A.~G.,  Semenov-Tian-Shansky M.~A.,
\newblock {\em Lax representation with a spectral parameter for the
 {K}owalewski top and its generalizations}. \newblock{Lett. Math. Phys.}, v.{14}, pp.55-61, (1987).

\bibitem{rs}
Reyman A.~G., Semenov-Tian-Shansky M.~A.,
 \newblock{ Integrable systems. The group-theoretic approach}, Institute for Computer Studies, Moscow-Izhevsk, 352 pp., (2003).

\bibitem{skl95}
 Sklyanin E. K.,
\newblock{\em Separation of variables-new trends}, Progr. Theoret. Phys. Suppl., v.118, pp. 35-60, (1995).

\bibitem{ts02}
Tsiganov A.V.,
	\newblock{\em On the Kowalevski-Goryachev-Chaplygin gyrostat},
J. Phys. A: Math. Gen.,  v.35,  L309-L318, (2002).


\bibitem{ts10}
Tsiganov A.V.,
\newblock{\em New variables of separation for particular case of the Kowalevski top},
Regul. Chaotic Dyn.,  v. 15, no. 6, pp. 657-667, (2010).

\bibitem{ts19s}
Tsiganov A.V.,
\newblock{\em Elliptic curve arithmetic and superintegrable systems},
{Physica Scripta}, v.94, 085207, 2019.

\bibitem{ts20r}
Tsiganov A.V.,
\newblock{\em Superintegrable systems and Riemann-Roch theorem},
Journal of Mathematical Physics, v.61, 012701, (2020).

\bibitem{ts20}
Tsiganov A.V.,
\newblock{\em Reduction of divisors for classical superintegrable $GL(3)$ magnetic chain },  	
Journal of Mathematical Physics, v.61, 112703, (2020).

\bibitem{ts21a}
Tsiganov A.V.,
\newblock{\em Reduction of divisors and Clebsch system},  preprint  	arXiv:2101.09993, 2021.	
preprint arXiv:2104.10362, (2021).

\bibitem{ts21}
Tsiganov A.V.,
\newblock{\em Reducible Abelian varieties and Lax matrices for  Euler's problem of two fixed centres },  	
preprint arXiv:2104.10362, (2021).

\bibitem{wei83}
Weil A.,
\newblock{\em  Euler and the Jacobians of Elliptic Curves}, In: Artin M., Tate J. (eds), Arithmetic and Geometry. Progress in Mathematics, vol 35. Birkh\"{a}user, Boston, 1983.

\end{thebibliography}
\end{document}